# Spatial resolution in atom probe tomography

Baptiste Gault[1,2], Michael P. Moody[1], Frederic de Geuser[3], Alex La Fontaine[1], Leigh T. Stephenson[1], Daniel Haley[1], Simon P. Ringer[1]

[1]*Australian Key Centre for Microscopy & Microanalysis, The University of Sydney, NSW, 2006, Australia.*

[2]*Department of Materials, University of Oxford, Parks road, Oxford, OX13PH, UK*

[3]*Science et Ingénierie des MAtériaux et Procédés(SIMAP) – UMR 5266 CNRS-Grenoble INP, Saint-Martin-d'Hères, France.*

Corresponding author: baptiste.gault@materials.ox.ac.uk

**Abstract**

This article addresses gaps in definitions and a lack of standard measurement techniques to assess the spatial resolution in atom probe tomography. This resolution is known to be anisotropic, being better in the depth than laterally. Generally the presence of atomic planes in the tomographic reconstruction is considered as being a sufficient proof of the quality of the spatial resolution of the instrument. Based on advanced spatial distribution maps, an analysis methodology that interrogates the local neighborhood of the atoms within the tomographic reconstruction, it is shown how both the in-depth and the lateral resolution can be quantified. The influences of the crystallography and the temperature are investigated, and models are proposed to explain the observed results. We demonstrate that the absolute value of resolution is specimen-specific.

# I. Introduction

Resolution, as it relates to microscopy, can be a difficult concept because it is complicated by an inherent reliance on arbitrary and/or subjective criteria. Further, even the meaning of the term 'resolution' can be a source of confusion. Commonly, its use refers to the *limit of resolution* which is defined as the minimum distance between two objects that will permit their distinction as two separate points within the image produced by the microscope. However, the term is also used in relation to the size of the image of a single point on the object, which is characterized by the pulse spread function (PSF) of the microscope. In regards to the limit of resolution, several criteria have been proposed as to how this distance can be best defined. The most commonly applied definition is Rayleigh's resolution criterion, an accepted standard in optics. The resolution itself is generally taken to be the full width half maximum of the pulse spread function. In the case of two distinct points, as they are brought together, the Rayleigh criterion leads to a limit of resolution is defined by the separation at which the PSF intensity at half the distance between two points is greater than about 80 % of the maximum intensity of a single spot (Van Dyck, et al., 2004).

The continual need to improve the resolution of atomic-scale imaging techniques has driven instrumentation development in a variety of fields of microscopy, including scanning probe microscopy, electron microscopy and atom probe tomography. Atom probe tomography is an atomic-scale microscopy and microanalysis technique that produces a chemically resolved three-dimensional image of the distribution of atoms within the bulk of a specimen (Blavette, et al., 1993a; Blavette, et al., 1993b; Cerezo, et al., 1988; Kelly, et al., 2004; Müller, et al., 1968; Nishikawa, et al., 2000). Recent major breakthroughs in the design of the instrument have enabled a significant broadening in both the field-of-view of the microscope as well as its field of applications. Since the resurgence of the pulsed laser atom probe (Bunton, et al., 2007; Cerezo, et al., 2007b; Gault, et al., 2006; Kellogg & Tsong, 1980; Tsong, et al., 1982), the technique has proven its potential as a metrology tool for both structural and functional materials, delivering an array of new insight for materials scientists (Blavette, et al., 1999; Cerezo, et al., 2007a; Cerezo, et al., 2007c; Miller & Kenik, 2004; Ringer, 2006; Seidman, 2007), especially correlated with atomistic materials modeling (Clouet, et al., 2006; Mao, et al., 2007). It is generally accepted that the spatial resolution in atom probe is anisotropic, being better in depth than laterally and, moreover, that the former is < 0.1 nm and the latter < 1 nm in the case of pure metals. We feel that there has been limited discussion of the actual physical origins of these resolutions and that clear definitions of these terms seem to be missing in the literature. Further several parameters are likely to impact the resolution, such as:

- the ellipticity of the specimen, which are often blade-shaped (Larson, et al., 1999),
- differences of behavior for different orientation for a single crystal (Chen & Seidman, 1971b),
- experimental conditions, including the temperature, pulse fraction, detection rate, the electric field,
- artefacts inherent to the field evaporation process itself such as roll-up, or field assisted surface diffusion (Tsong & Kellogg, 1975; Waugh, et al., 1976),
- trajectory aberrations (Vurpillot, et al., 2000a) and local magnification (Miller & Hetherington, 1991) linked to structural or, in the case of alloys, chemical variations within the specimen, that may also affect the sequence of evaporation (De Geuser, et al., 2007),
- and the reconstruction algorithm used for positioning the atom.

Addressing some of these issues is the primary purpose of this paper in the ideal case of a pure metal, in order to develop the general understanding of the physical phenomena underpinning the spatial performance of atom probe tomography.

In other fields of microscopy, the development of an aberration-corrected environment in TEM has necessitated new standard definitions for assessing microscope resolution (Batson, et al., 2002; Haider, et al., 1998). The characteristic dumb-bell shapes that can be observed in Si along the [110] direction provide a convenient metric since the 0.136 nm spacing between the atoms is defined with a high degree of precision and is difficult to image unless the point resolution of the microscope is below 1 Å. Similarly, single Au atoms, nanostructures or surfaces have also often been used. The resolution is generally derived from the presence of the corresponding spatial frequencies in the Fourier transform of the image and is then described as an information limit. Recently, sub-0.5 Å information limit was achieved on the Transmission Electron Aberration-corrected Microscope (TEAM) based on an assessment of Au [110] surfaces as a benchmark for the measurement (Kisielowski, et al., 2008). This approach to estimate the resolution in atom probe tomography had previously been proposed by Kelly *et al.*(Kelly, et al., 2007), however, it would only enable assessment of the information limit, not the actual spatial resolution.

The present article proposes standard procedures to assess the spatial resolution in atom probe tomography based on the application of spatial distribution maps (SDMs) (Geiser, et al., 2007) as applied to a pure aluminum specimen, as well as a modeling of the different aspects of the physics underpinning the variation of the resolution. Aluminum provides an excellent test case for the analysis as it is readily available to any researcher, is easily prepared, specimen are rarely blade-shaped and crystallographic features are always present in the reconstruction facilitating accurate calibration of the tomographic reconstruction (Gault, et al., 2008; Gault, et al., 2009b).

## II. Basics of atom probe tomography

As depicted in Figure 1, in atom probe tomography, the atoms from the surface of a needle-shaped specimen are ionized and desorbed from the surface under the effect of a very intense electric field, in a process known as field evaporation(Müller, 1956). The electric field is partly pulsed to enable the chemical characterization of each detected ion by time-of-flight mass spectrometry. This process generally takes place nearly atom-by-atom, and layer after layer, in a very specific order, that is related to the distribution of the electric field at the surface. The dimensions and shape of the specimen make the electric field highly divergent in the vicinity of the tip apex, providing a very high magnification of up to several millions to the corresponding projection. This enables the resolution of individual atomic positions, similarly to that observed in field ion microscopy, the first technique enabling the observation of atomically resolved images of surfaces (Müller, 1957; Müller, 1965). Once an ion is generated, it is thus projected, under the action of the electric field, onto a single particle position-sensitive detector.

Unlike electron microscopy, the detection of a given species does not relate to the Z number of the atom, but is a statistical process directly linked to the use of micro-channel plates as electron amplifier at the entry of the detector. Subsequent to the experiment a tomographic reconstruction of the probed volume is built atom-by-atom by means of an inverse point-projection algorithm. This approach to the 3D reconstruction developed by Bas *et al.* (Bas, et al., 1995) is summarized in Figure 2. The in-depth coordinate of every atom is based upon the sequence in which they arrived at the detector. Each atom simply contributes a successive increment proportional to its atomic volume to the in-depth extent of the reconstruction. Consequently an additional corrective term applied to the in depth position of each atom to account for the curvature of the original surface. The Bas et al. procedure was designed for instruments with small fields-of-view, and therefore many approximations assuming low angles in the projection are no longer valid in the case of modern large field-of-view atom probes. Consequently corrective terms are added to account for the increased field-of-view (Geiser, et al., 2009). Despite its intrinsic simplicity, this reconstruction procedure enables highly accurate reconstruction, at least in the ideal case of pure metals.

Different ways to interrogate a typical example of tomographic reconstruction of a pure aluminum data set is displayed in Figure 3. A top view reveals the pole structure (a), while a side view shows several atomic plane families directly imaged by the technique (b). The result of advanced spatial distribution maps analyses, and of a Fourier transform on the left part (Vurpillot, et al., 2001; Vurpillot, et al., 2004b) are both displayed, respectively in (c) and (d). Spatial distribution maps (SDMs) are an atom probe tomography data mining technique that reveal the average local neighborhood of an atom, and can be used to discern specific crystallographic features or structure present in the reconstruction. Recent research has focused on the development of SDM techniques enabling the user to tune in to the location and crystallographic orientation in which of the local atomic neighborhood is scanned within the analyzed reconstruction (Moody, et al., 2009). The spatial distribution maps are generally displayed, as shown in Figure 3, either as a z-SDM, a 1D plot revealing the average atomic interspacing in the in-depth direction of a specific crystallographic orientation, or as a *xy*-SDM, a 2D histogram highlighting the average local arrangement of the atoms within the lateral plane.

## III. Resolution in atom probe tomography

The spatial resolution in atom probe tomography is highly anisotropic: the resolution in the direction of the analysis, the dimension that is built through the procedure described above, is often high enough to resolve atomic planes in the depth of the material (Warren, et al., 1998), as revealed in Figure 3. However, the lateral resolution is more limited, and structures in the plane perpendicular to the direction of analysis appear blurred (Vurpillot, et al., 2000b; Vurpillot, et al., 1999). Based on Fourier transform calculations, Vurpillot *et al.* defined the spatial resolution as the width of the damping function that limits the intensity of the peaks in the reciprocal space (Vurpillot, et al., 2001). Using a combination of Fourier transforms, they were able to measure the in-depth and lateral resolution. Their approach was however based on the assumption that the resolution along a direction was a combination of a resolution normal and one resolution parallel to the analyzed surface. Their results were then confronted with predicted results from a model built derived from field ion microscopy and only valid for rare gas atoms surrounding a charged surface (Chen & Seidman, 1971a; Tsong, 1990). The validity of this approach seems thus questionable. More recently, Kelly *et al.* proposed a definition based on the image analysis of a thin slice of a spatial distribution map. The spatial frequencies within the image were revealed using Fourier transform, and they defined the spatial resolution as the longest vector of the reciprocal space (Kelly, et al., 2007). This definition is somehow erroneous, this approach reveals down to what interspacing the instrument can image atomic planes, which is the definition of the *limit of information* and not the resolution in itself.

Lateral resolution is thought to be limited by a combination of trajectory aberrations in the early stages of the flight of the ion (Vurpillot, et al., 2000a; Vurpillot, et al., 2000b; Waugh, et al., 1976) and thermal agitation of the atoms at the surface that provides lateral velocity to the ions, slightly deflecting their trajectory. Inextricably linked to the reconstruction procedure, the parameters impacting the in-depth resolution are less simple to explain. Since the in-depth coordinate is attributed to the atoms via sequential increments, a change in the sequence in which the atoms are evaporated from the surface should affect the in-depth dimension. However, no actual quantification has ever been performed to support these assertions.

In Figure 3 advanced spatial distribution maps have been computed in a pure aluminum data set. The results are displayed as a 1D distribution of atomic inter-distances along the in-depth direction (*z*-SDM) enabling accurate measurement of the planar interspacing, and also as a 2D histogram highlighting the average local arrangement of atoms within a plane normal to a crystallographic direction (*xy*-SDM). Similar to the definition proposed by Vurpillot *et al.* for reciprocal space analysis, the width of a single peak in the SDM analyses reflects the local spatial resolution in the respective real space directions. Therefore it is possible to quantify the lateral and in-depth resolution via post-processing measurements. This approach can be seen as assuming that the pulse spread function of the atom probe as a microscope is a Gaussian functions, the standard deviation of which reflects either the in-depth or the lateral resolution.

## IV. Experimental

Specimens in the experiments presented here were prepared from high purity (99.999%) aluminum, using standard electrochemical polishing using solutions of 10% perchloric acid in butoxyethanol at 5-15V. They were subsequently analyzed using two different Imago LEAP-3000X Si equipped with 8 cm delay line detectors at a pulse fraction of either 0.15 at 33K ± 2 K and 92mm flight path, 0.25 with a specimen temperature varied between 20 and 200 ± 2 K and a flight length of 90mm, and various pulse fractions ranging from 15 to 35% at 20 ± 2 K. All experiments were conducted under ultrahigh vacuum conditions (<4.5 10−9 Pa). The average detection rate was controlled at to obtain an average of 0.04 ions per pulse for the first, 0.025 for the second, and 0.035 for the last experiments. For each experiment, between 2 and 7 million atoms were collected to ensure statistics, and to make sure the specimen is studied in an equilibrium state. Indeed, it is known that the specimen shape evolves to accommodate the temperature and field conditions (Drechsler, 1992 ; Drechsler & Wolf, 1958; Webber, et al., 1979). This shaping requires atoms to be field evaporated, and we assumed that the first million of atoms collected in each condition might not reflect the actual shape of the specimen, but a transient state during which the specimen is

## V. In-depth resolution

### *Influence of the crystallography*

First advanced $z$-SDM were computed for several atomic plane families within the same data set. Consequently, for each $z$-SDM, the central peak of the distribution has been isolated, and fitted to a Gaussian function, as depicted in Figure 4. The resolution can then be defined as $\delta = 2\sigma$, with σ the standard deviation of the fitted Gaussian function. This definition is a counterpart in the real space of the one given by Vurpillot *et al.* in the reciprocal space (Vurpillot, et al., 2001). A plot of δ as a function of its relative lattice interspacing reveals a linear relationship (Figure 5). We have recently revealed this unexpected trend in a letter (Gault, et al., 2009a).

## *Spatial distribution of the resolution*

It is also worth mentioning another interesting methodological development enabled by the use of advanced spatial distribution maps (Moody, et al., 2009) and the determination of the in-depth resolution presented in the present paper. Advanced SDMs permit the scanning of the local distribution of the atoms in depth along a specific orientation and around a specific location. As revealed by Figure 3, an atom probe tomographic reconstruction often exhibits a succession of different atomic planes families. The resolution of planes in a particular crystallographic directions is best in a small cross-section region of interest (ROI) centered on the corresponding pole. By systematically moving the ROI, in which the advanced SDM is calculated, away from the centre of the pole, the evolution of the in-depth resolution along this direction can be assessed, as shown in Figure 6. Due to the curvature of the specimen surface, a given set of plane can be imaged until another set of planes with a different orientation starts to be imaged. The surface can thus be seen as a succession of facets, each of those facets corresponding to a crystallographic direction. If this structure is easily observed in field ion microscopy, we show here that this can be also derived from the atom probe tomography data: when the ROI is moved away from the pole, the quality of the planes in this direction quickly deteriorates, revealing the extent of the facet for a crystallographic direction.

## *Experimental dependence of in-depth resolution with pulse fraction*

A pure aluminum specimen was analyzed at 20 K across a series of varying pulse fraction. The values of the resolution obtained on the [0 0 2] pole are reported in Figure 7 (a). The resolution does not show any significant change with a change in the pulse fraction in a reasonably large range of pulse fraction commonly used in atom probe tomography. Inspection of the charge state ratio reveals that the electric field conditions are exactly similar for all the different experiments. Desorption maps do not show significant change in the distribution of the local density of collected atoms. Changes in the density can be directly linked to local changes in the magnification that is proportional to the curvature of the specimen. Therefore significant changes in desorption maps would reveal changes in the actual shape of the specimen. Here, the shape global shape of the specimen is not significantly affected by the variations in the pulse fraction. The trend to an apparent improvement of the resolution with the pulse fraction is consistent with the progressive increase in the radius of curvature from the first to the last experiment,.

## *Experimental dependence of in-depth resolution with temperature*

The very same aluminum specimen was analyzed at several different temperatures, and the resolution was estimated for three poles over the specimen surface. The resulting evolution of the resolution is plotted in Figure 7 (b). Unexpectedly, this graph shows that, despite affecting the sequence of evaporation, an increase in temperature up to 80K does not dramatically affect the in-depth resolution. This indicates that systematic errors introduced by the reconstruction procedure are prominent over any effect induced by the temperature. Above a critical temperature though, the resolution quickly worsens.

## *Discussion*

To explain the apparent dependence of in-depth resolution on crystallography and temperature, let us consider a simple evaporation model in which a single atomic terrace represents the entire the analyzed area (Figure 8 (a)). Let us consider that the atoms in this model are evaporated in an ideal order, at first from the periphery of the terrace, known as kink-site atoms, then progressively inwards and are detected with a perfect accuracy. Atoms in-depth positions should approximate a line with a thickness of a single atomic in-depth increment, resulting in an extremely thin peak in the $z$-SDM, as shown in Figure 8 (b). However, in reality as the positions of arrival of the atoms on the detector are blurred by a combination of the positioning accuracy of the detector and the finite lateral resolution, the curvature correction induces a broadening of the width of this plane, as for Figure 8 (c). This leads to a constant term for a given position on a given specimen.

The specimen temperature should also play a role. Indeed, it can be considered that, since field evaporation is a probabilistic thermally assisted process, temperature can induce some degree of randomness in the sequence of evaporation of atoms from the tip. This means that not only atoms from the outermost edges of a terrace on the surface are evaporated. Evaporation of non-kink site atoms has been observed experimentally observed by various authors (Moore & Spink, 1969; Stiller & Andren, 1982). Each field evaporated non-kink site atom adds an in-depth increment to the width of the PSF. This phenomenon is thermally activated and accounts for the dependence of the in-depth resolution with the temperature. This will degrade the resolution in the reconstruction as atoms will not be attributed the correct number of in-depth increments corresponding to their actual depth as illustrated in Figure 8 (d).

The degradation of the in-depth resolution is thus intimately linked to the geometrical size of a terrace, which directly relates to the tip radius and to the interspacing of the considered atomic plane family (Drechsler & Wolf, 1958; Gault, et al., 2008; Miller, et al., 1996). The number of non-kink evaporated atoms from a given terrace can be written as the number *N* of atoms on the terrace multiplied by Maxwell-Boltzmann probability. To each of these events corresponds to a broadening of the PSF from the in-depth increment *dz*. The temperature-dependent resolution broadening term can finally be written as $N\, dz \exp\left(-\frac{\Delta E}{k_B T}\right)$, where ΔE is a measurement of how much stronger the non-kink atoms are bonded to the surface than the kink atoms. Considering the formula given by Bas et al.(Bas, et al., 1995), the in-depth increment is given by $dz = V_{at}/\eta\, S_a$, while the number of atoms on the terrace can be related to the size of the terrace and the atomic volume: $N = \frac{S_a\, d_{hkl}\, \eta}{V_{at}}$. In these formulae, η is the detector efficiency. Finally, the in-depth resolution can be written as the sum of a constant term depending on purely geometrical considerations (crystallography and radius of curvature of the specimen) and a temperature dependant term: $\delta = \delta_0 + d_{hkl} \exp\left(-\frac{\Delta E}{k_B T}\right)$. This relation is in good agreement with the linear relationship of the in-depth resolution with $d_{hkl}$ observed in Figure 5. This agreement is a good indication that there is no complex dependence of $\delta_0$ with $d_{hkl}$. From the expression of d, the difference between the evolution of the resolution in the 2 experiments shown of Fig. 5 can be understood. The difference in the slope arises from the difference of analysis temperature while the difference in the straight lines offsets is a consequence of a different radius of the tip. A dimensionless parameter can subsequently be extracted from the previous expression: $\Delta = \frac{\delta}{d_{hkl}} = \frac{\delta_0}{d_{hkl}} + \exp\left(-\frac{\Delta E}{k_B T}\right)$. The dimensionless resolution, Δ, can be used as a more general criterion, independent of the specific more general criterion which do not depend on the specific crystallography. It is worth noting that the value of this dimensionless resolution depends on the radius of the tip, and is thus neither constant over the course of an experiment nor for a given material. It is strictly dependant on the considered specimen. This is the reason underpinning the differences in the values achieved in the two different experiments shown in Figure 5. Based on the post-ionization theory (Haydock & Kingham, 1980; Kingham, 1982), the electric field in the vicinity of the specimen surface was deduced from the charge state ratio of aluminum, and it is found that it is around 2% higher in the case of the specimen analyzed at 40K (diamond symbols), which results in a change in the slope. The change in the slope between the different experiments can be attributed to the change in the field conditions. The average energy barrier $\Delta E$ in one case and the other is 5.4 meV (40K) in the first case and 5.8meV in the second one (33K).

The formula presented here for the variation of the in-depth resolution is also in good agreement with the observed dependence of the resolution on the temperature. Below 100K, the term $\delta_0/d_{hkl}$ is prominent over the temperature dependant term. However, as the temperature increases, a fast degradation of the resolution ought to be expected due to the exponential nature of the probabilistic phenomenon associated with the corruption of the evaporation sequence, which is in very good agreement with our experimental observations. Further, the fact that not all the poles are affected in the same way can be explained, first, by the specific evaporation field of the different facets, as discussed in previous work (Chen & Seidman, 1971b; Miller, et al., 1996). The influence of the shape of the tip, which is dramatically affected by a change in the temperature above 100K as we have previously shown in a previous article (Gault, et al., 2009b), can also play an important role and may be responsible for the discrepancy in the evolution of the resolution for the [1 1 3] and [0 0 2] poles at 140K.

Finally, based on the expression derived above for $\Delta$, and using the parameters directly extracted from a linear fit of the experimental values presented in Figure 5, we can estimate the limit of resolution as a function of the temperature. This limit of resolution $d_{hkl}^{(\lim)}$ represents the shortest interspacing that can be resolved in a tomographic reconstruction. Considering limited variations of $\delta_0$ and $\Delta E$ with the temperature, the limit of resolution can be estimated as a function of the temperature as shown in Figure 9. These curves also relate on the choice of $\Delta_{lim}$ a parameter that defines the limit of resolution. Assuming a Gaussian pulse-spread function applied to a set of atomic planes, a first value of $\Delta$ was derived from the definition of the Rayleigh criterion (dotted line). An experimental criterion was also estimated, based on *z-SDM* analyses of reconstructions generated from the same specimen across a series of experiments at several temperatures. A significantly larger value of $\Delta=0.715$ was obtained, indicating that the Rayleigh criterion is extremely conservative. The curve defines the atomic plane imaging space: to be imaged a set of planes has to have an interspacing larger than the value of the plotted curve. This indicates that the reconstruction procedure is very robust, and that large changes in the experimental conditions, and therefore specimen shape, will not significantly affect the tomographic reconstruction, until other physical phenomenon start being prominent over the standard field evaporation of kink-site atoms.

## VI. Lateral resolution: results and discussion

Atom probe specimens are typically cooled down to cryogenic temperature to avoid electric field-induced surface diffusion and hence supposedly to minimize the loss in spatial resolution. The lateral resolution in field ion microscopy had been the subject of numerous studies. Müller *et al.*

(Müller, et al., 1965), presented the basis of a model, in which the gas atoms were assumed to migrate towards the apex of the specimen via a series of leaps above the surface. Every time a gas atom bounced over the surface of the tip, it would release some of its thermal energy, hence progressively reducing its velocity. Chen and Seidman (Chen & Seidman, 1971a) subsequently developed a more sophisticated model based on the assumption of perfect gas atoms ionized in the vicinity of the specimen apex, gave relatively accurate numerical estimations of the resolution. However, even though aspects of field ion microscopy and atom probe tomography are similar, the main assumptions of these models are not valid in the specific case of atom probe tomography.

*Limit of resolution as a function of crystallography*

At the surface, the atoms are oscillating within a potential energy well, the lateral dimensions and depth of which strongly depend on the packing of the atoms over the considered plane, and hence specific facets on the surface of specimen. This is demonstrated in Figure 10, in which the inter-atomic potentials at the surface of the model tip, described above, were plotted for two different crystallographic orientations ([0 0 2] and [1 1 5]). The shape of the potential restricts the extent of movement on the surface within the lateral plane induced by thermal agitation of the atoms. This movement considered partly responsible for the limited lateral resolution of the technique, which can be thought of as representative the average space occupied by a single atom oscillating inside the limits of the potential well binding the atom to the surface. Other effects originating from trajectory aberrations (Miller, 1987; Vurpillot, et al., 2000a; Vurpillot, et al., 2000b; Vurpillot, et al., 2004a) or roll-up effect (Schmidt & Ernst, 1992; Waugh, et al., 1976) also induce a degradation of the lateral resolution. However, it is difficult to deconvolute one effect from the others.

To reveal the average two-dimensional local arrangement of atoms within the identified crystallographic planes, a series of advanced spatial distribution maps were computed (Moody, et al., 2009). A radial distribution function was then applied to the *xy*-SDM, as displayed in Figure 11. Where possible Gaussian functions have been fitted to the observed peaks. The first peak (in green) corresponds to the interaction between an atom and its first shell of nearest neighbors. It is proposed that the width of this first peak is a direct representation of the lateral resolution, $\lambda$, of the technique ($\lambda=2\sigma$ of the Gaussian function) in this particular crystallographic direction.

To quantify how the spatial resolution is impacted by the local atomic structure within the plane or by thermal agitation, this technique was used to determine the lateral resolution across several crystallographic directions, and plotted as a function of in-depth interspacing, $d_{hkl}$, at a constant temperature of 40K, in Figure 12. This graph shows a trend to an improved lateral resolution as the interspacing of the corresponding set of planes progressively increases. The degree of packing within a given plane is proportional to the two-dimensional atomic density. Assuming that the density of the material is isotropic, we can assume that the packing within a plane is inversely proportional to the corresponding interspacing between planes in this direction. Subsequently the data points were used to fit a function $\lambda = \lambda_0 + \frac{s_0}{d_{hkl}}$, where $\lambda_0$ represents the contribution of trajectory aberrations, and $s_0$ describes the average surface occupied by an atom over the specimen surface due to thermal agitation and is homogeneous to a surface. It must be noted that the reconstruction procedure used in atom probe tomography does not consider atoms as spheres, but as points without volume. Therefore $s_0$ is the average surface equivalent to this point oscillating in a potential well at a given temperature. The fitting parameters give $\lambda_0 = 104.3 \; pm$ and $s_0 = 4319 \; pm^2$.

## *Limit of resolution as a function of temperature*

The procedure was then repeated several temperatures for the analysis of the atomic distribution within the [002] planes, as presented in Figure 13, which shows an expected trend of an increase of the lateral resolution with increasing temperature. A second variation law of the lateral resolution, now as a function of the temperature can be derived. It can be assumed that the resolution will have a constant component due to the trajectory aberrations $\lambda_0$, independent of temperature. A second term relates to the probability for an atom to move on the surface within a distance corresponding to the position of its first nearest neighbor. An offset of the atom by more than this distance would imply a complete loss of determinism, since it would be impossible then to discriminate between an atom and its first neighbor. Based on these assumptions, we can write $\lambda = \lambda_0 + d_{1NN} \exp\left(-\dfrac{\Delta E_{migration}}{k_B T}\right)$ where $d_{1NN}$ is the distance to the first nearest neighbor, $k_B$ is the Boltzmann constant, $T$ the absolute temperature and $\Delta E_{migration}$ is the energy barrier for the atom to migrate. This displacement could take place laterally of by roll-up motion to a meta-stable position close to the atomic site. The latter possibility has been experimentally observed before (Schmidt & Ernst, 1992). The fitting parameters are $\lambda_0 = 109.7 \; pm$ and $\Delta E_{migration} = 10.2 \; meV$.

The contribution of trajectory aberrations can thus be assumed to be around 100 pm for this analysis, despite changes in the radius of curvature of the specimen and changes in the mesoscopic scale shape of the specimen with temperature (i.e. ref. (Gault, et al., 2009b)). This strongly indicates that the influence of trajectory aberrations are effectively linked to the very local neighborhood of the atoms (Moore, 1981). Further, the temperature for which the resolution would be of the order of the distance to the first nearest neighbor extracted from the fitting function, $\approx 243\,K$, is in very good agreement with experimental observations, since the very same specimen analyzed at 230K, displayed in Figure 13, showed a structure of strongly indicative of surface migration. Finally it is worth noting the very good agreement between the values of $\lambda_0$ obtained from both fits, all these results tends to validate our approach for investigation of the lateral resolution in atom probe tomography.

## VII. Conclusion

In conclusion, this study defines new standard for the assessment of the resolution in atom probe tomography.

- Methods to quantify both the in-depth and lateral resolutions have been proposed based on the use of advanced spatial distribution maps available to most atom probe users.
- For the first time the resolution in the atom probe reconstruction has been directly related to the crystallography of the specimen, highlighting the importance of the geometry and local topology of the specimen surface.
- Further, we have investigated the influence of pulse fraction temperature on the spatial resolution.

- Underpinning this investigation, we have shown that the in-depth resolution's dependence on both crystallography and temperature can be directly linked to their influence on the sequence of evaporation in the experiment.
- The physics behind these variations has been discussed using simple theoretical models developed to support experimental observations.

Continued research will see these models progressively evolve to higher complexities that permit an understanding of the evolution of the spatial resolution in the case of multi-component materials and non-metallic materials. Several parameters that are likely to impact the resolution have not been explored yet. In particular, the impact of the chemistry of the material or the ellipticity of the specimen on the reconstruction are still to be explored. This first step towards improving the general understanding of these phenomena is the pathway to improve the accuracy of the reconstruction procedure in atom probe tomography.

**Acknowledgements**

The authors would like to thank Drs Ross Marceau and Tim Petersen for fruitful discussions. The authors are grateful for funding support from the Australian Research Council, which partly sponsored this work. The authors are grateful for scientific and technical input and support from the Australian Microscopy & Microanalysis Research Facility (AMMRF) at The University of Sydney.

**Figure Captions**

Figure 1: Schematic view of the local electrode atom probe, XD, YD are the coordinates of the impact of the ion on the position sensitive detector, N refers to the position of atom within the sequence of detected atoms.

Figure 2: Diagram of the sequential procedure used to build the tomographic volume: (1) the atoms are successively field evaporated from the specimen in a given order, (2) each atom contributes to the depth by increment that directly relates to its atomic volume, (3) a correction term is computed for each atom to account for the curvature of the emitting surface, (4) final reconstruction showing the atomic planes structure (not to scale in depth), (5) corresponding intensity of the spatial-distribution map assuming a Gaussian pulse-spread function.

Figure 3: Different way to interrogate an atom probe microscopy pure aluminum analysis at 40 K (a): slice in the tomographic reconstruction encompassing the {0 0 2}, {1 1 5}, {1 1 3}, {1 1 1} atomic planes (b). Spatial distribution maps in depth ($z$-SDM) along the {2 0 10} direction (c). Fourier transform shows traces of high harmonics relating to several atomic plane families, and highlights the limit of information of the technique (d). A lateral ($xy$-SDM) along the {1 1 1} direction (5) enable to obtain average information along these directions.

Figure 4: Spatial Distribution Map in z (z-SDM) oriented along the {1 1 5} planes, showing the sequence of planes and isolated central peak of the z-SDM fitted with a Gaussian function, enabling the measurement of the resolution.

Figure 5: Plot of the resolution as a function of the interspacing $d_{hkl}$ along a given crystallographic direction hkl for pure Aluminum specimens, one with a radius ~ 63 nm analyzed at 40K and .025 ions per pulse (square) and one with a radius of ~ 73 nm analyzed at 33K and 0.04 ions per pulse (diamonds).

Figure 6: **(a)** Change in the resolution as the advanced SDM is scanned around the 113 pole, here along the x axis. A Gaussian function has been fitted to the experimental data and is shown as a guide for the eyes. Reproducing this procedure along several directions enables to dress a two-dimensional distribution of the resolution along this direction (b), which correspond to the region in which the 113 planes are imaged.

Figure 7: Dimensionless resolution as a function of the pulse fraction **(a)** and temperature for three poles **(b)**.

Figure 8: Model atom probe reconstruction: the colors correspond to atoms having the same probability to be field evaporated (a). Reconstruction procedure applied to this terrace: first, in the case of a perfect sequence of evaporation (b); then, in the case of a finite lateral-resolution, the curvature correction induces errors in the reconstruction (c); and finally, considering that the sequence of evaporation can be affected by the probabilistic nature of the field evaporation phenomenon (d).

Figure 9: Plot of the limit of resolution $d^{(lim)}_{hkl}$ as a function of the temperature.

Figure 10: Shapes of the potentials binding the atoms on a facet of the tip corresponding to a 002 plane (1) and a 115 plane (2). Please note the difference in the packing and the thus the area in which the atoms are free to vibrate on one facet and the other.

Figure 11: Radial Distribution Function computed from a *xy*-SDM corresponding to the 002 planes, in which Gaussian functions have been fitted to successive peaks. The first fitted peak directly reflects the lateral resolution of the technique and its width was used as a measurement of this resolution. The others peaks are just guides for the eyes, since boundary effects and blurring effect make the other peaks overlapping.

Figure 12: Lateral resolution as a function of the interspacing (circles) and fit (dashed line).

Figure 13: Lateral resolution as a function of the temperature (squares) and fit (dashed line). The thick horizontal dotted line corresponds to the distance to the first nearest neighbor.

Figure 14: Desorption map of the pure Al specimen presented in figure 3, analyzed at 230K.

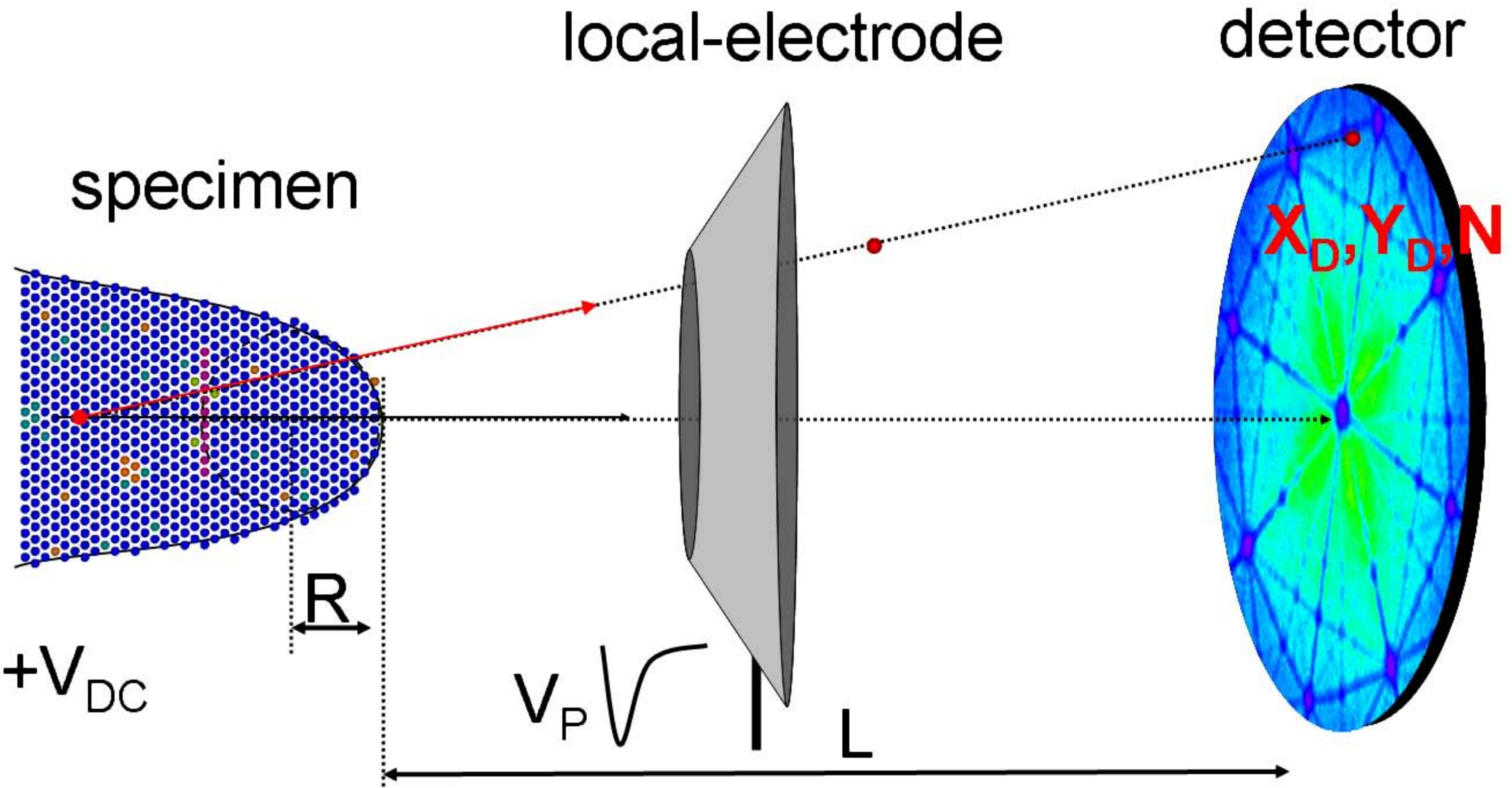

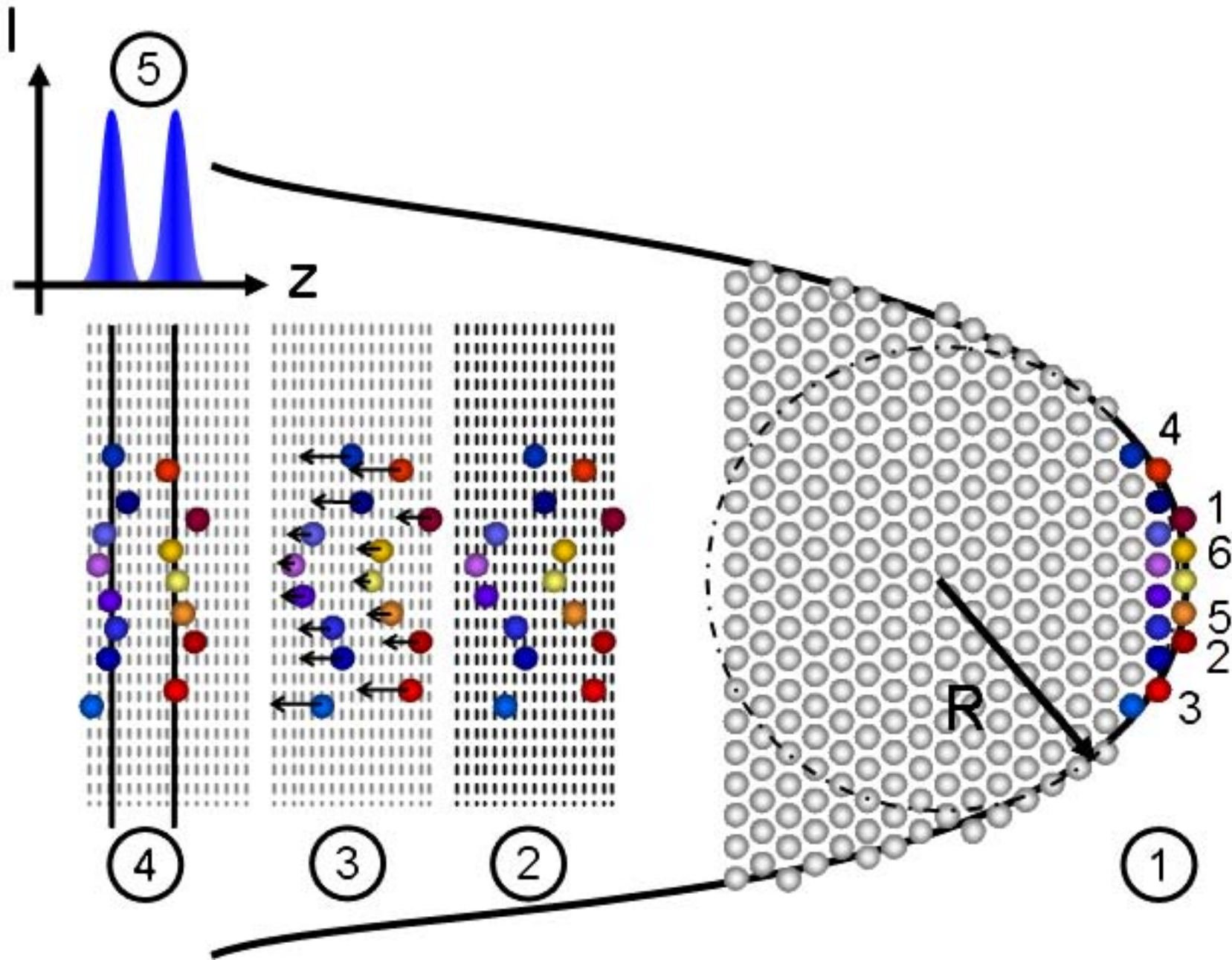

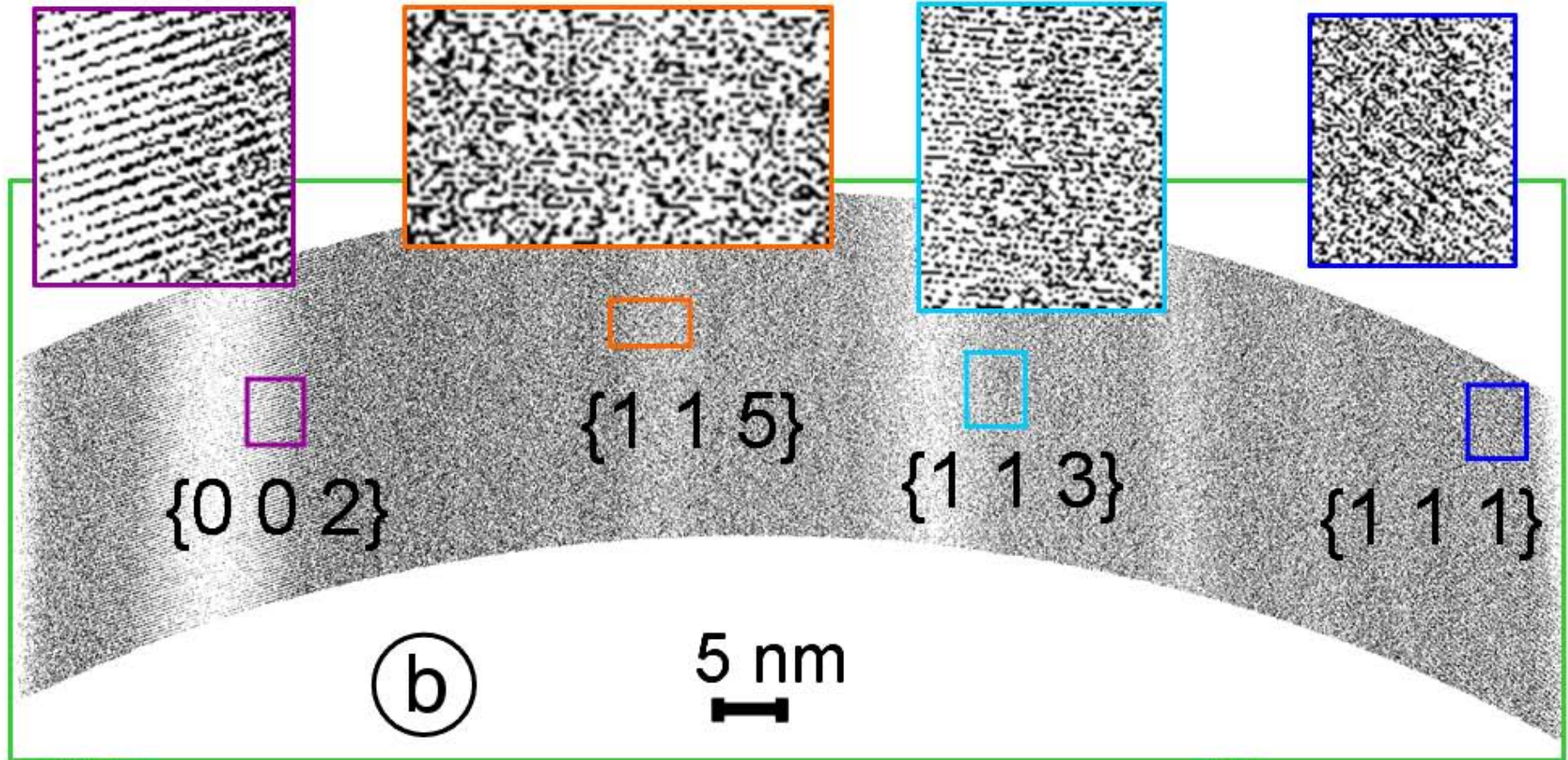
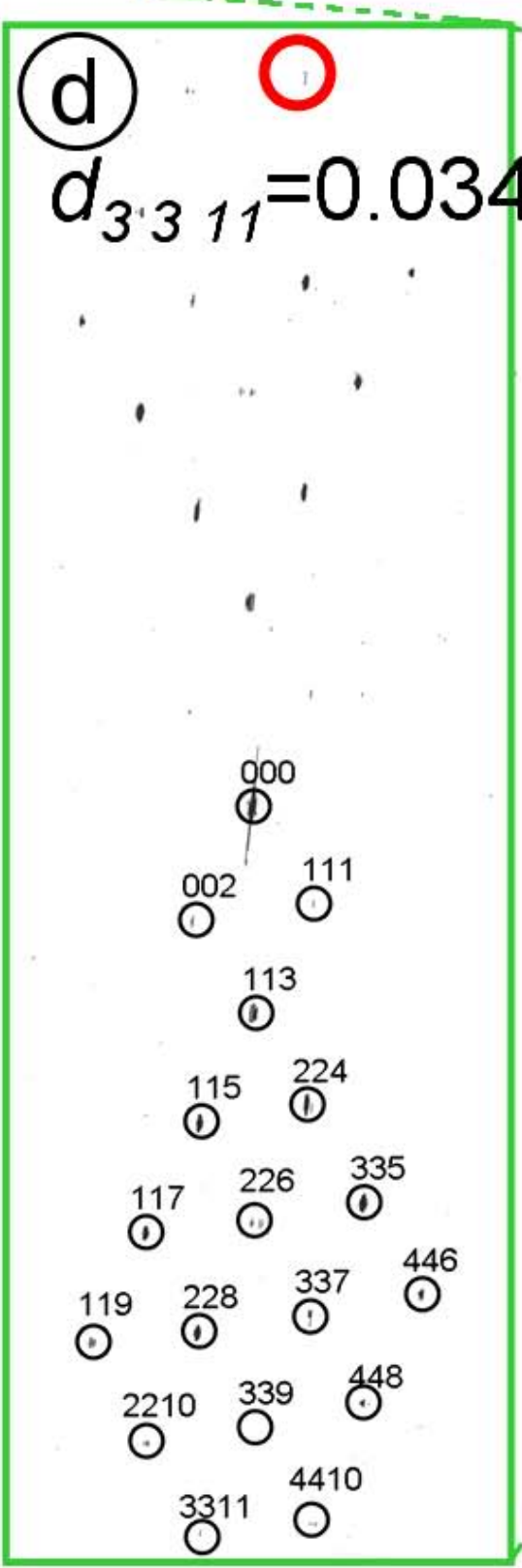
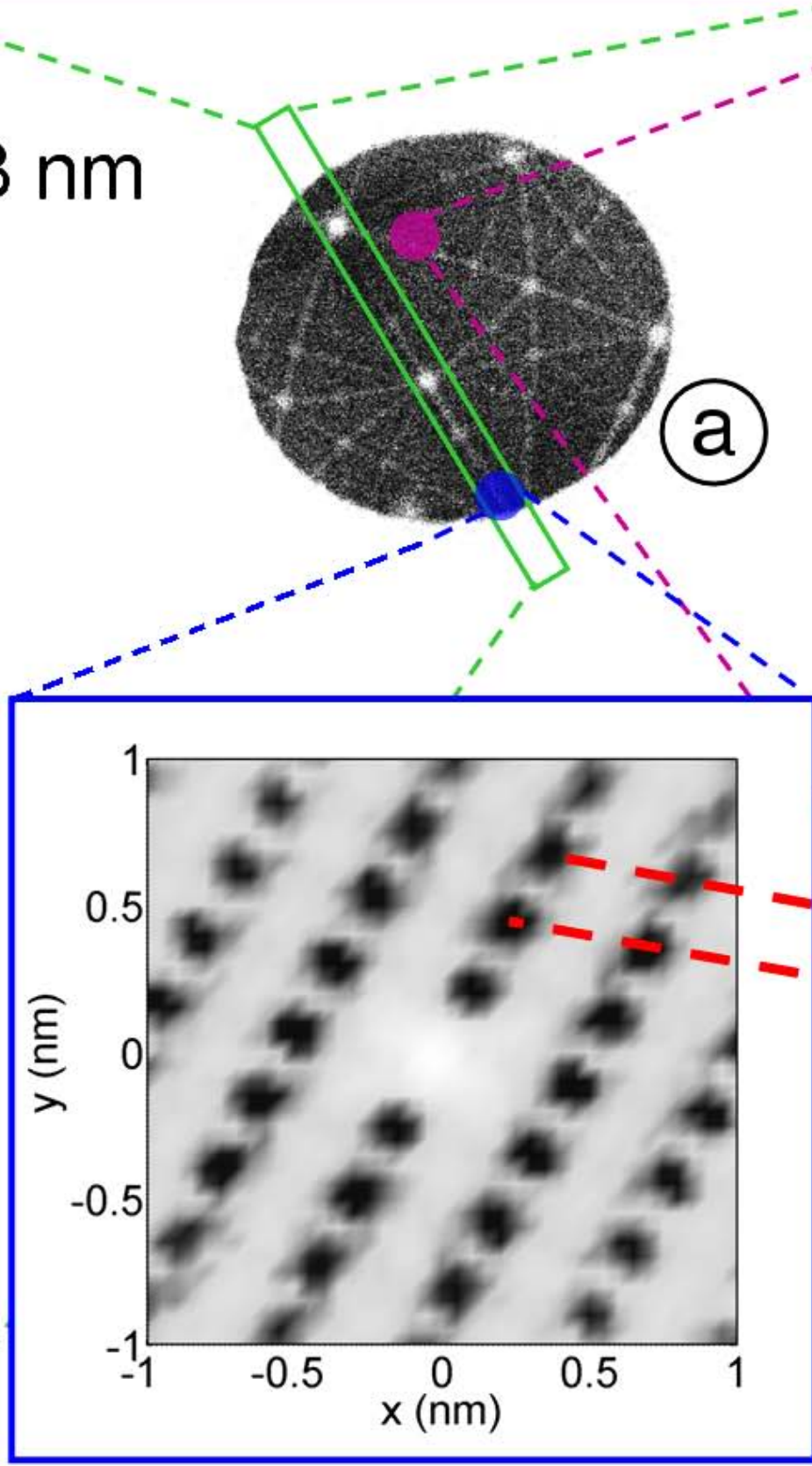
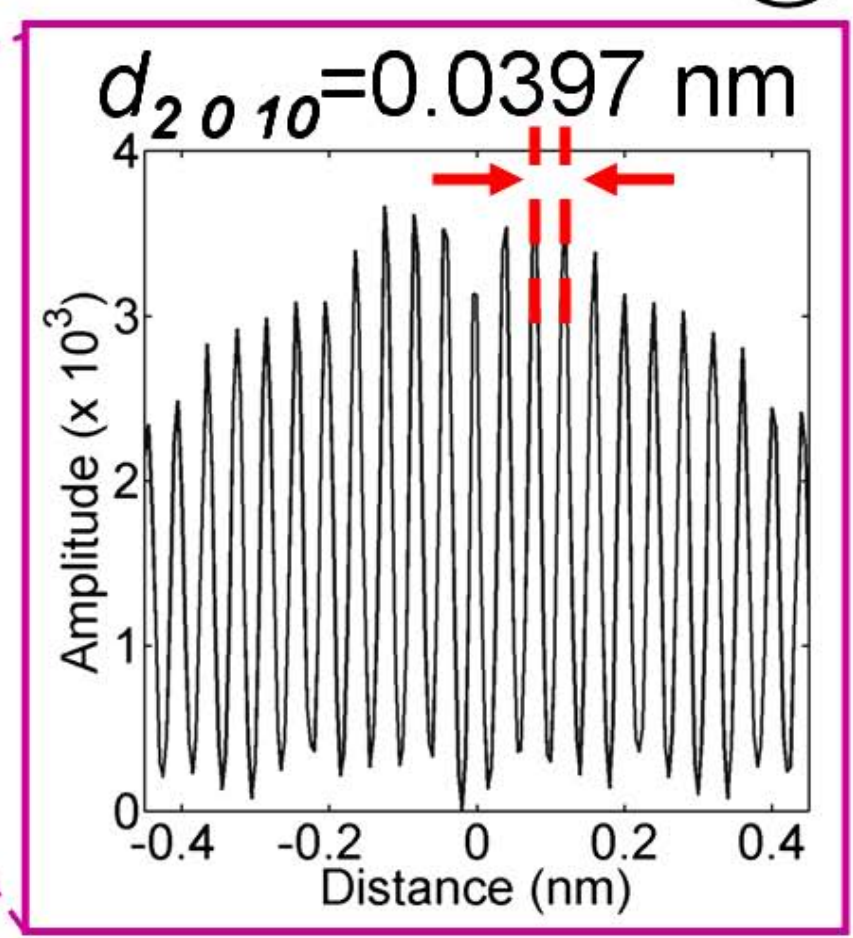
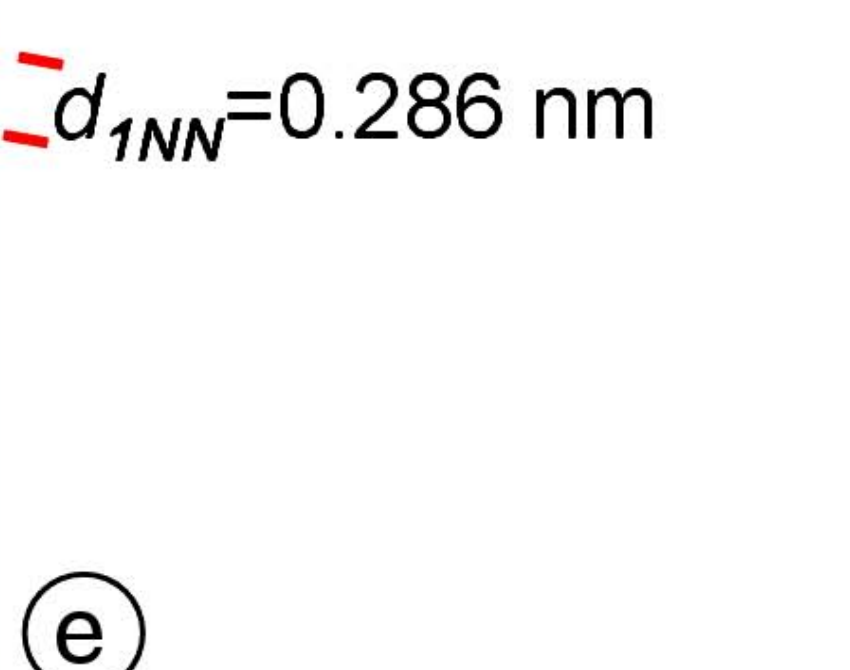

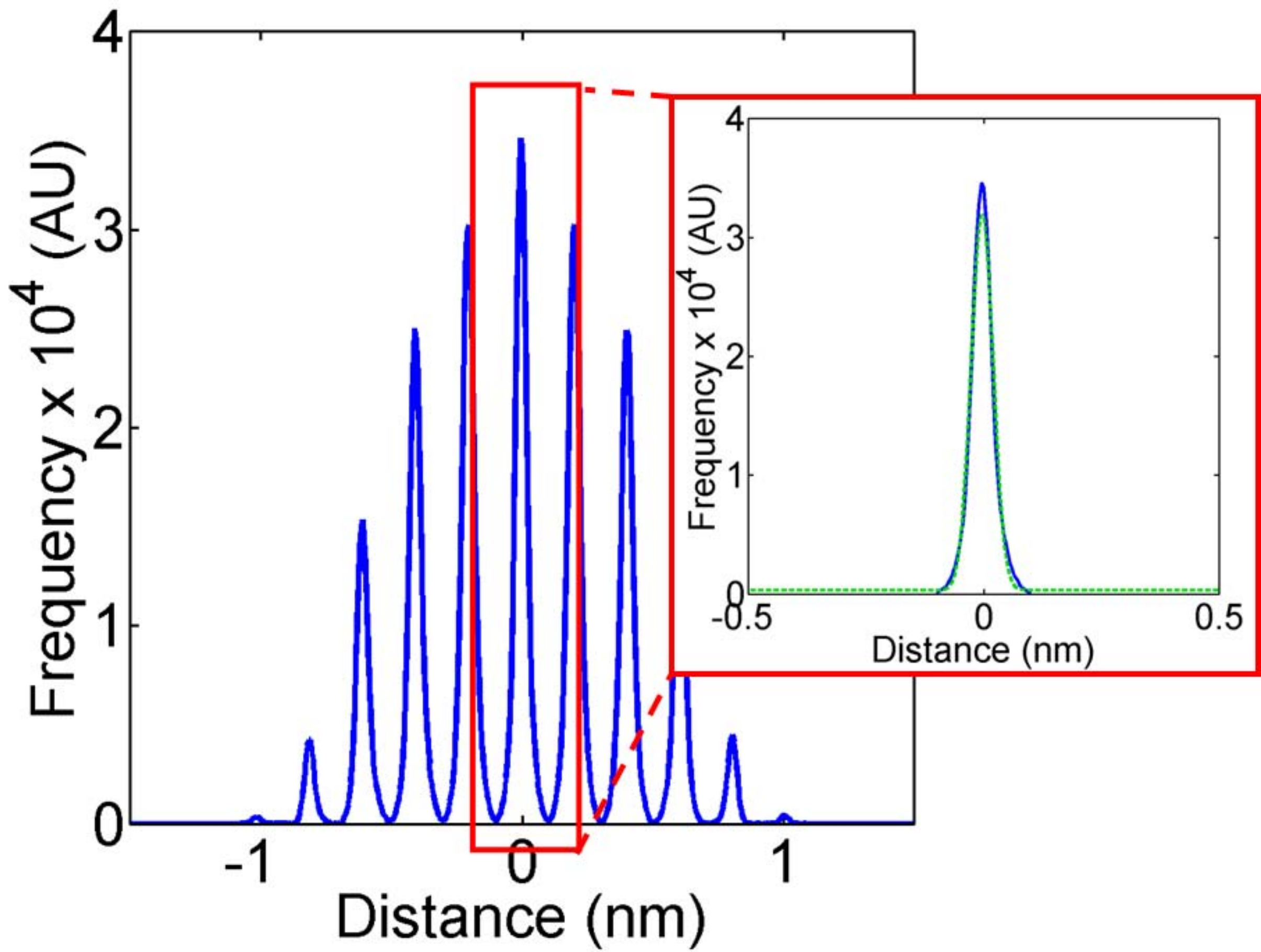

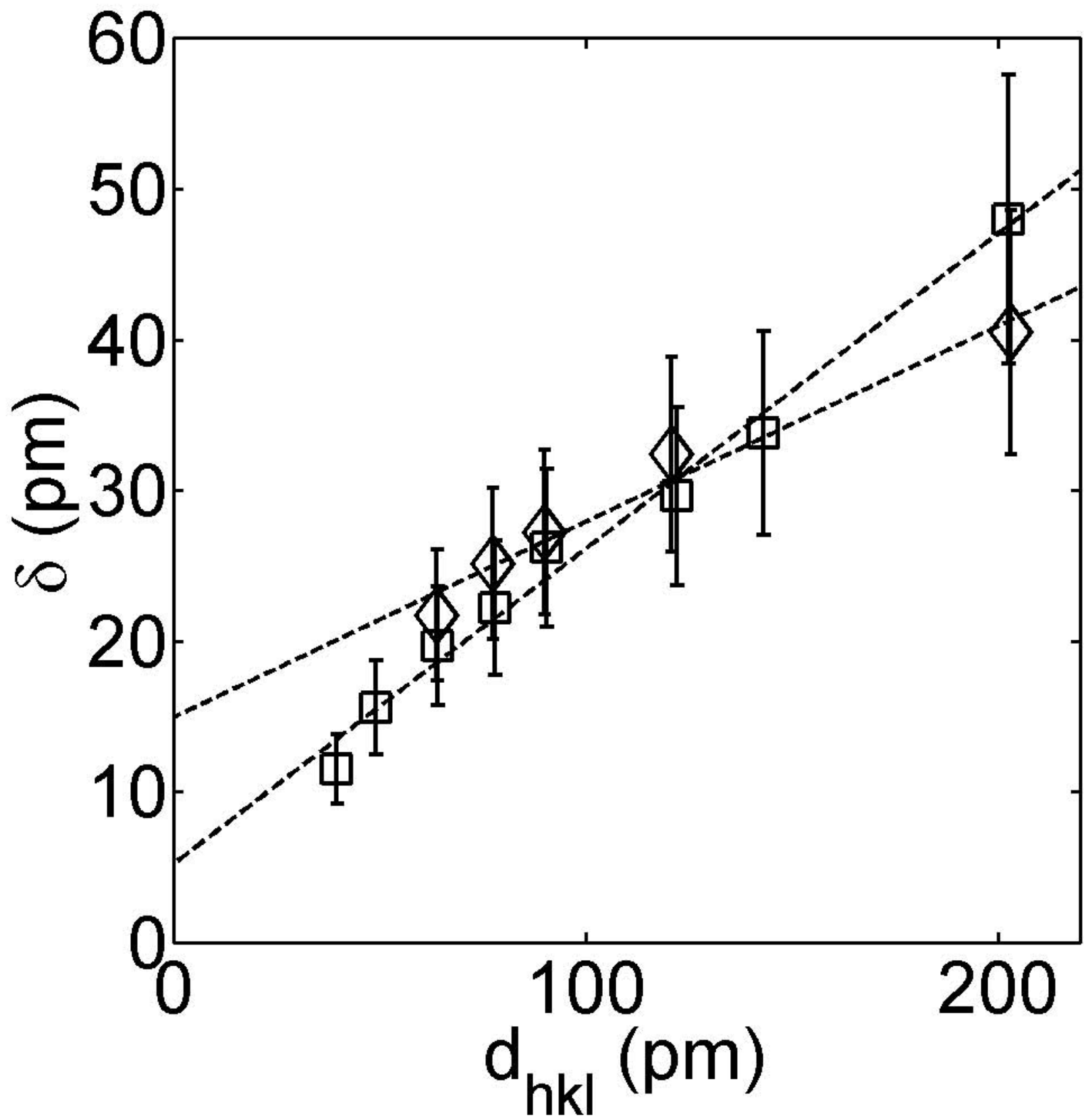

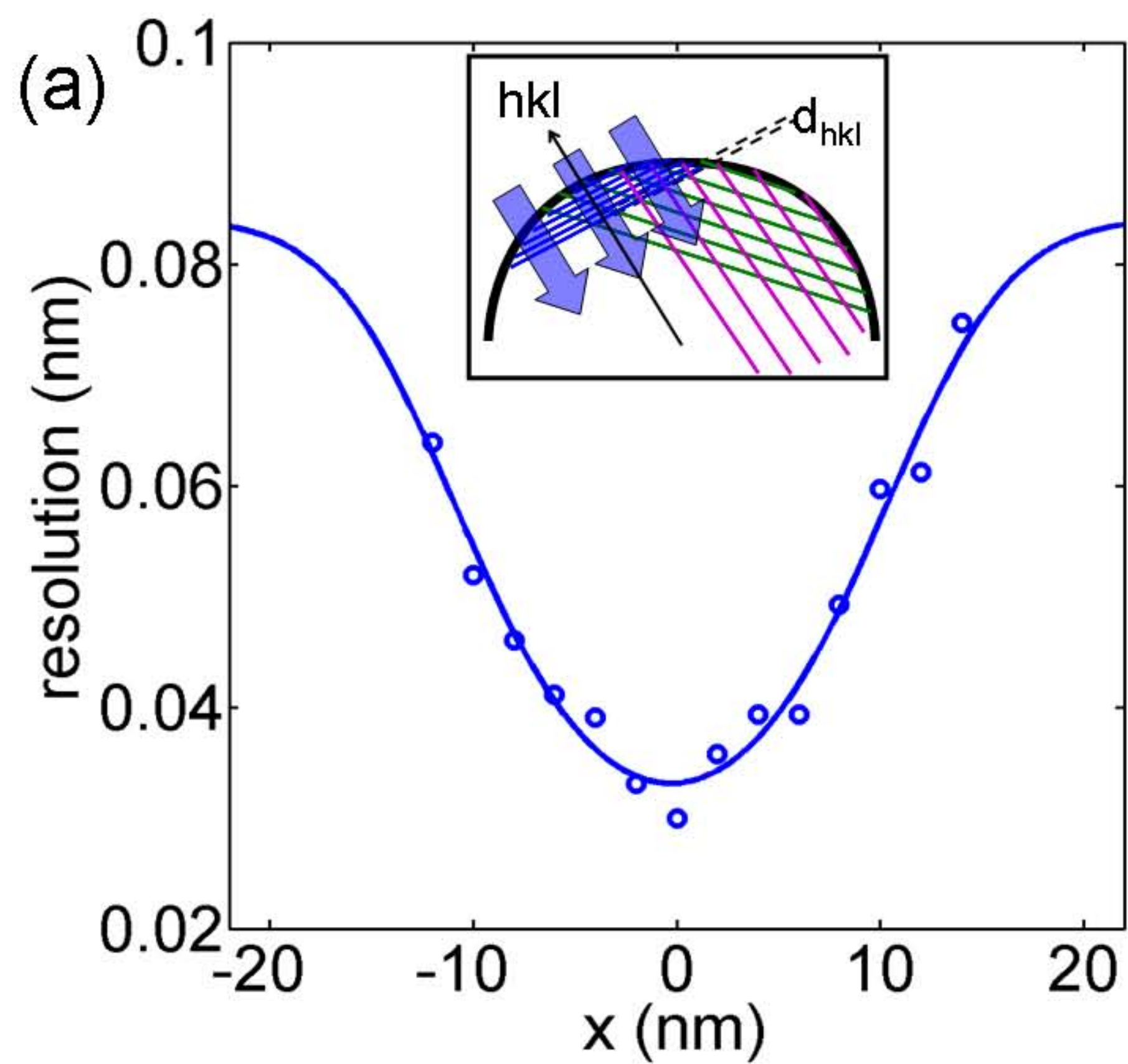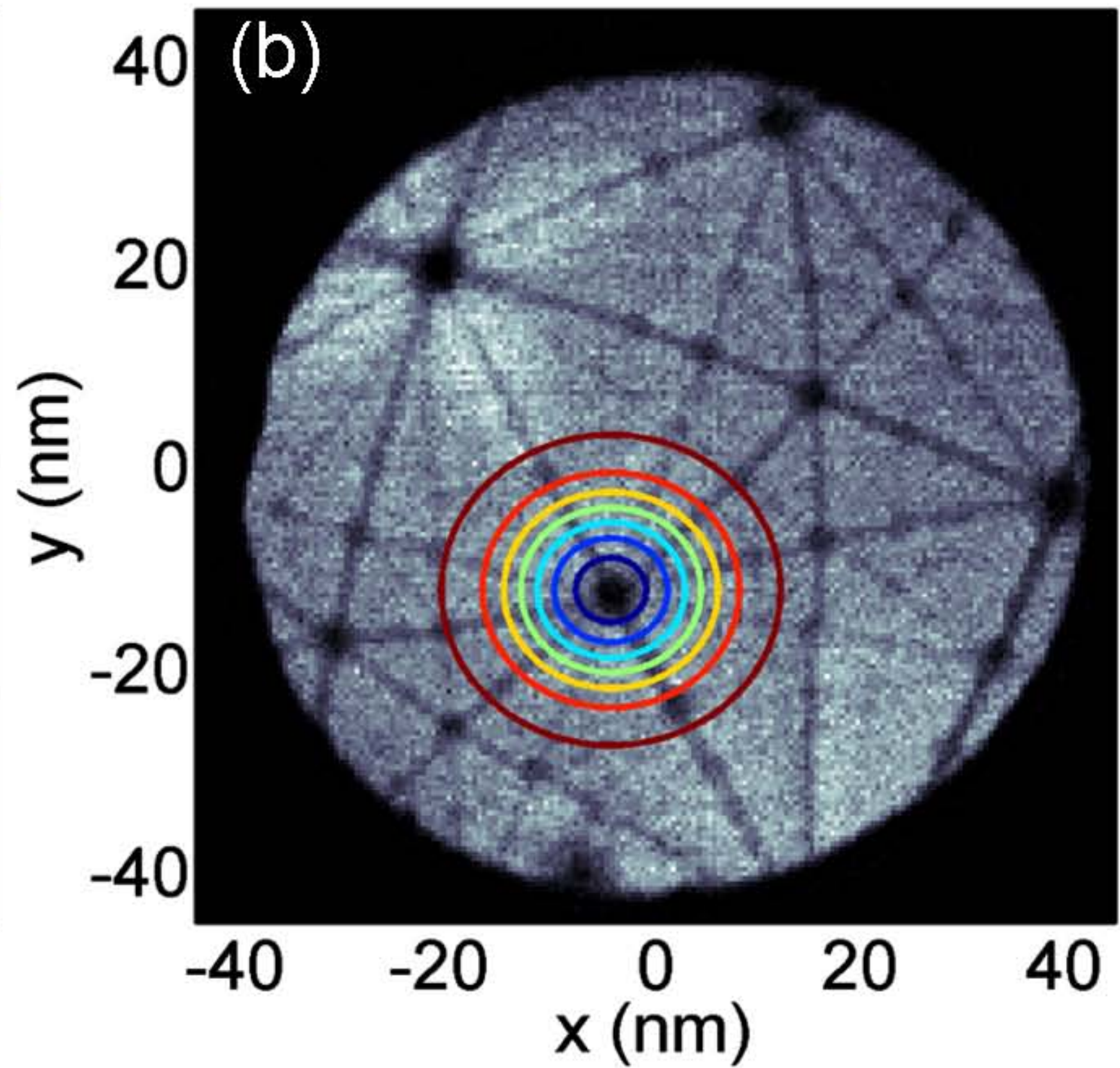

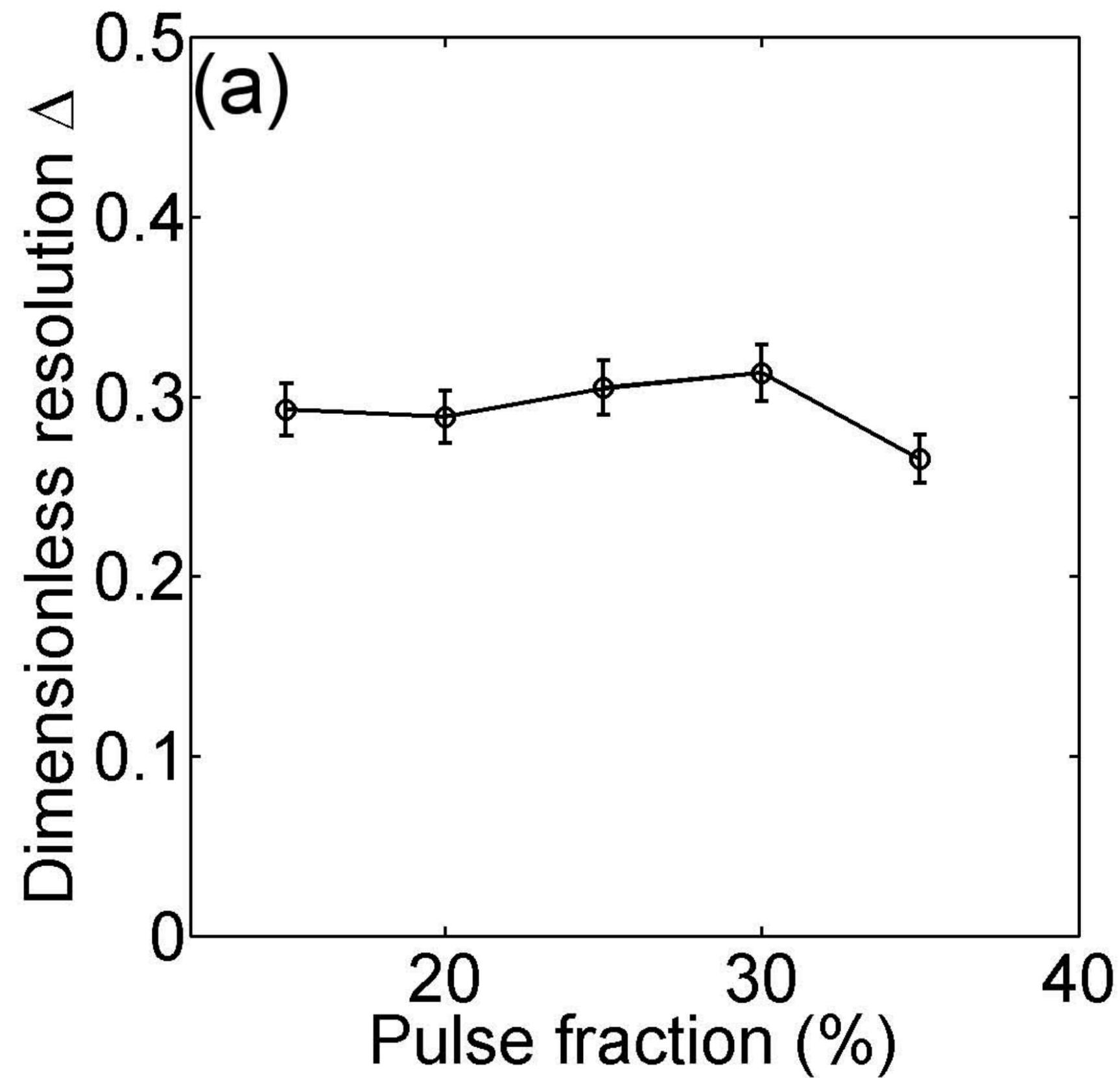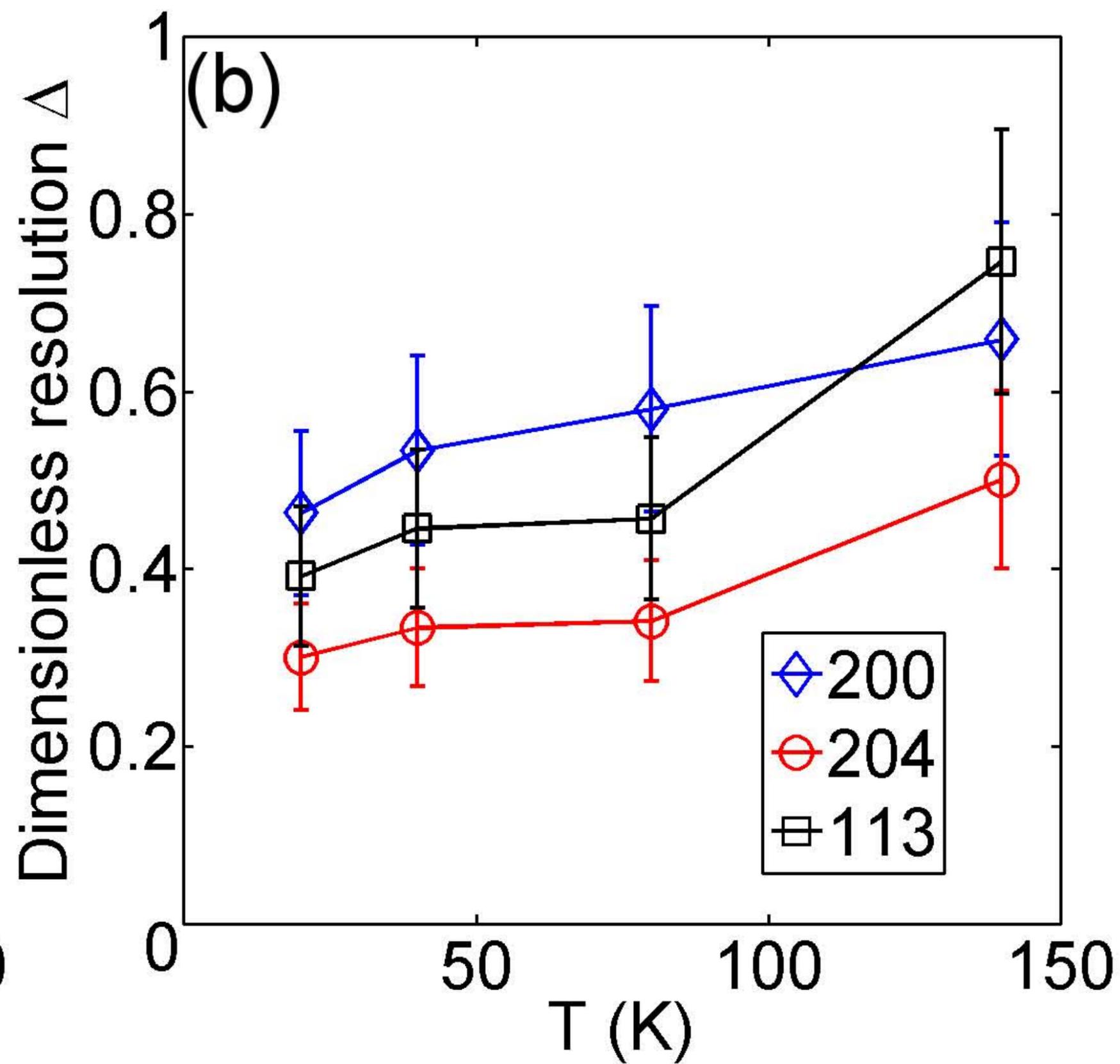

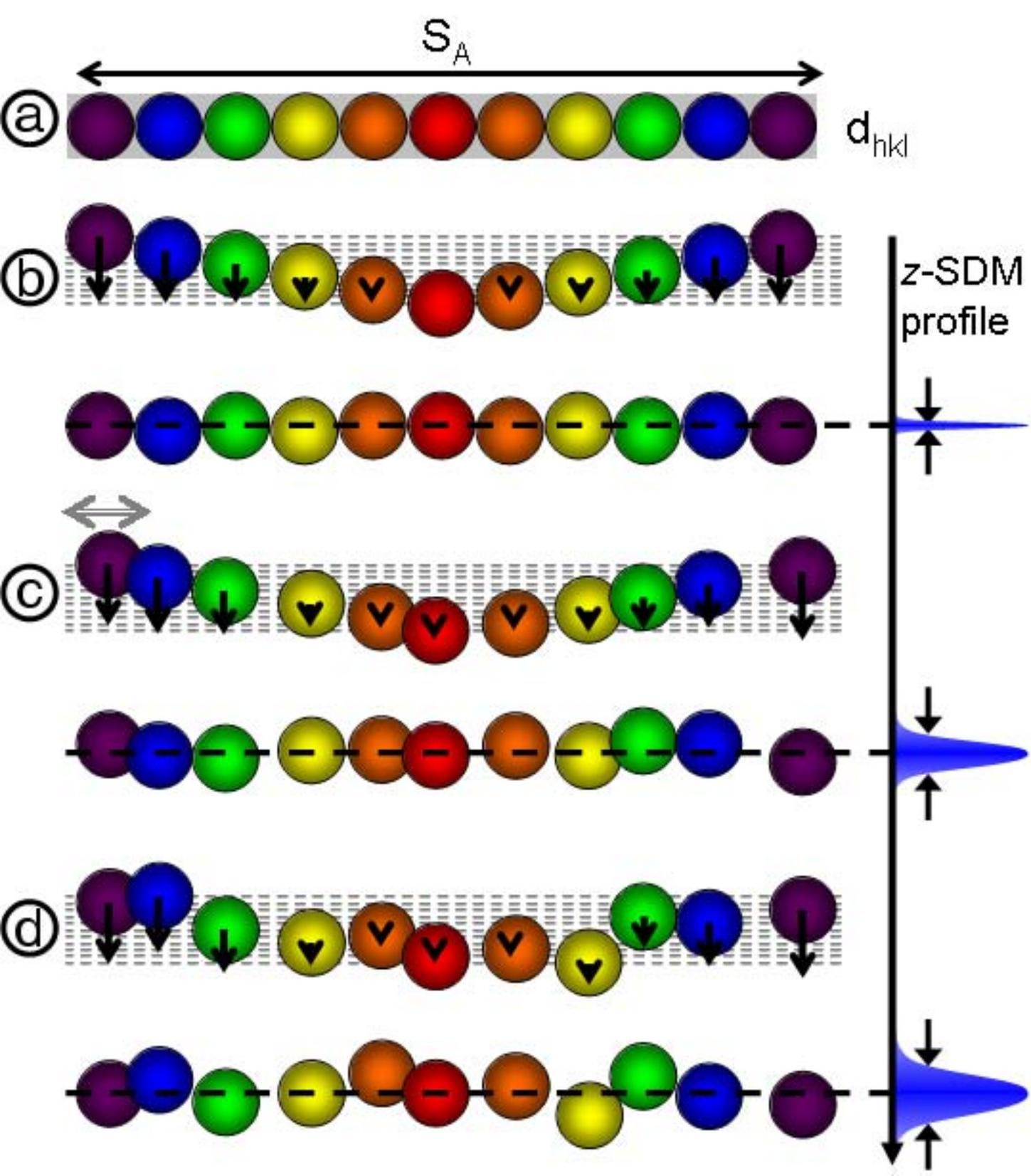

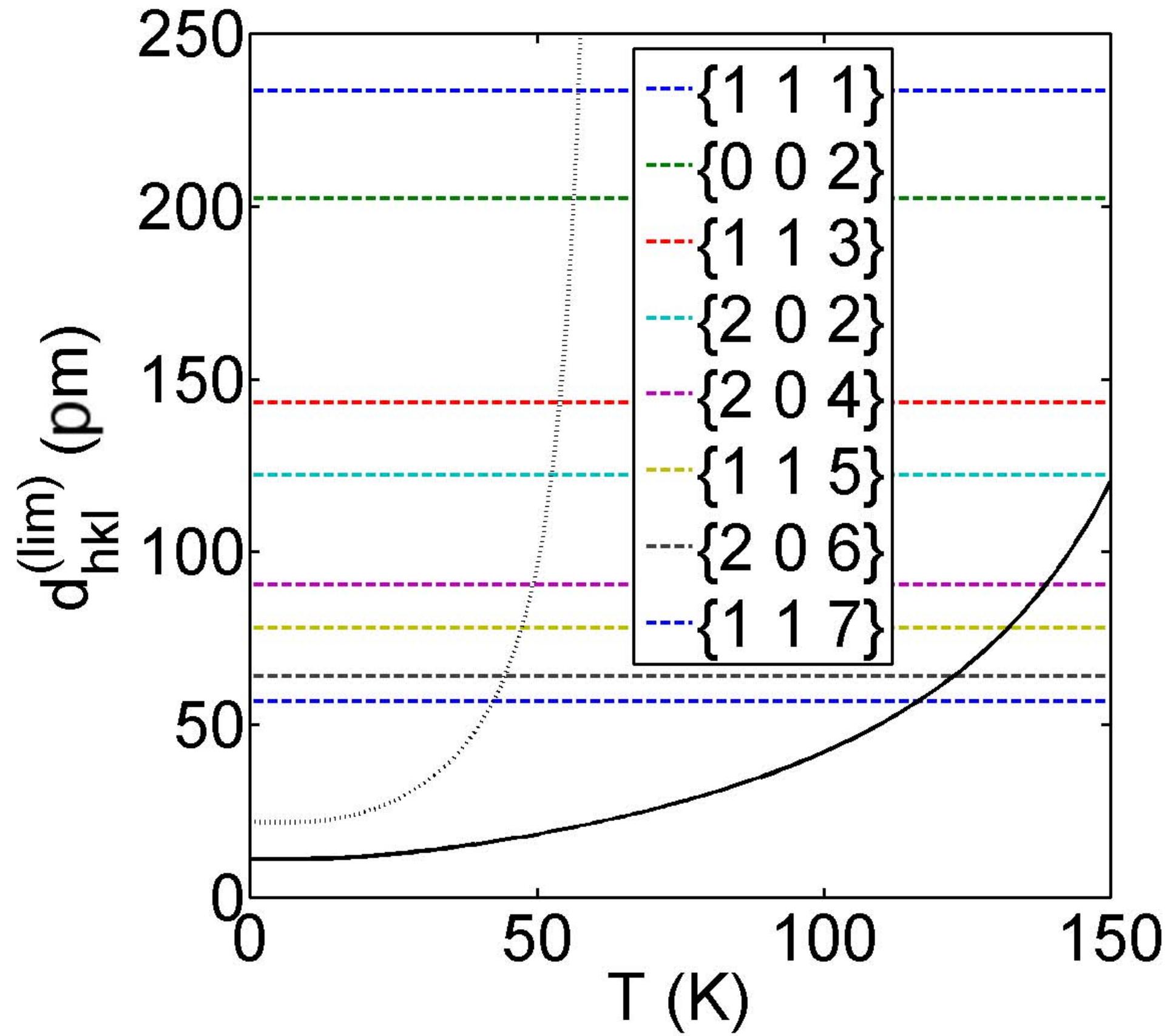

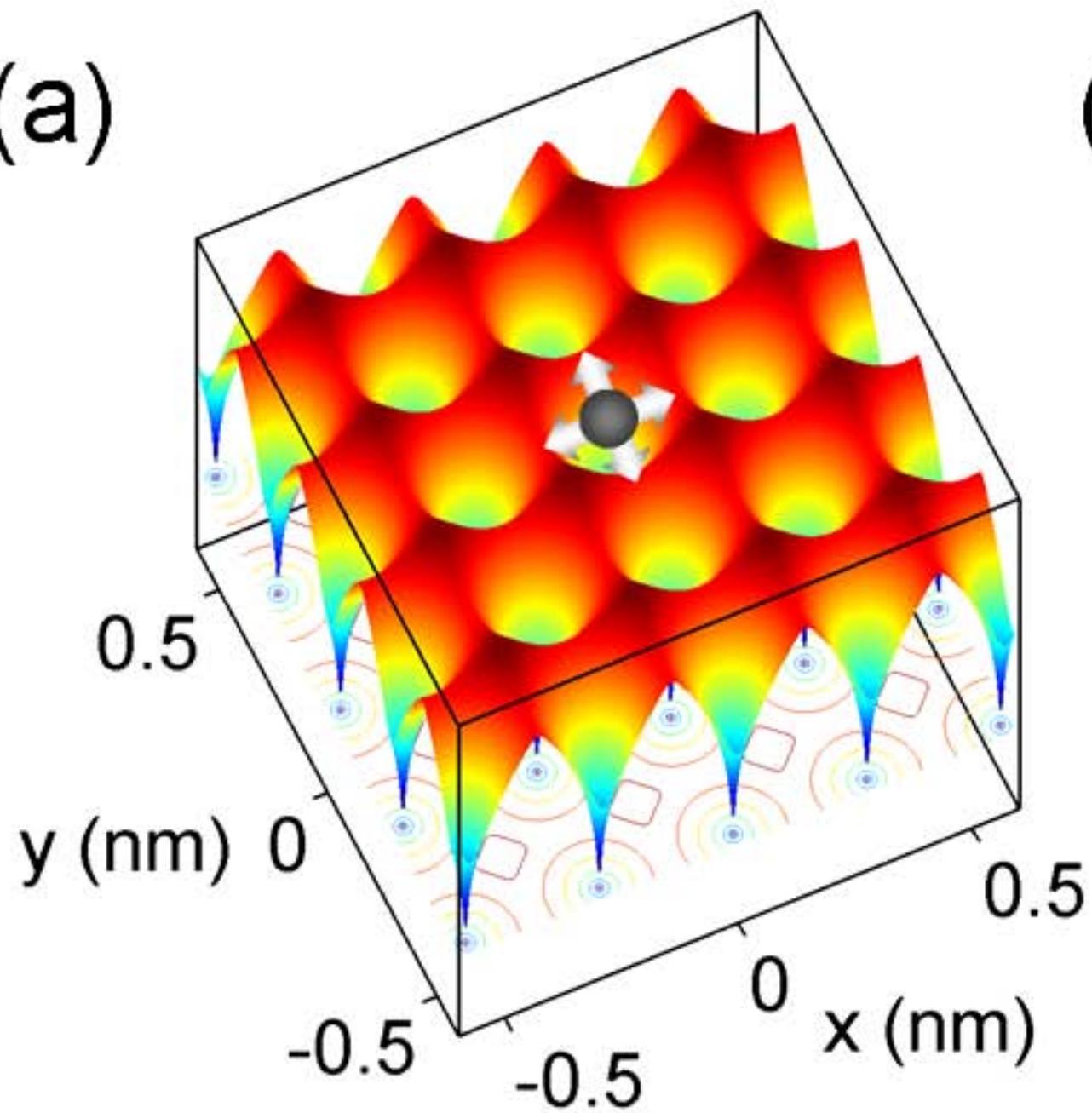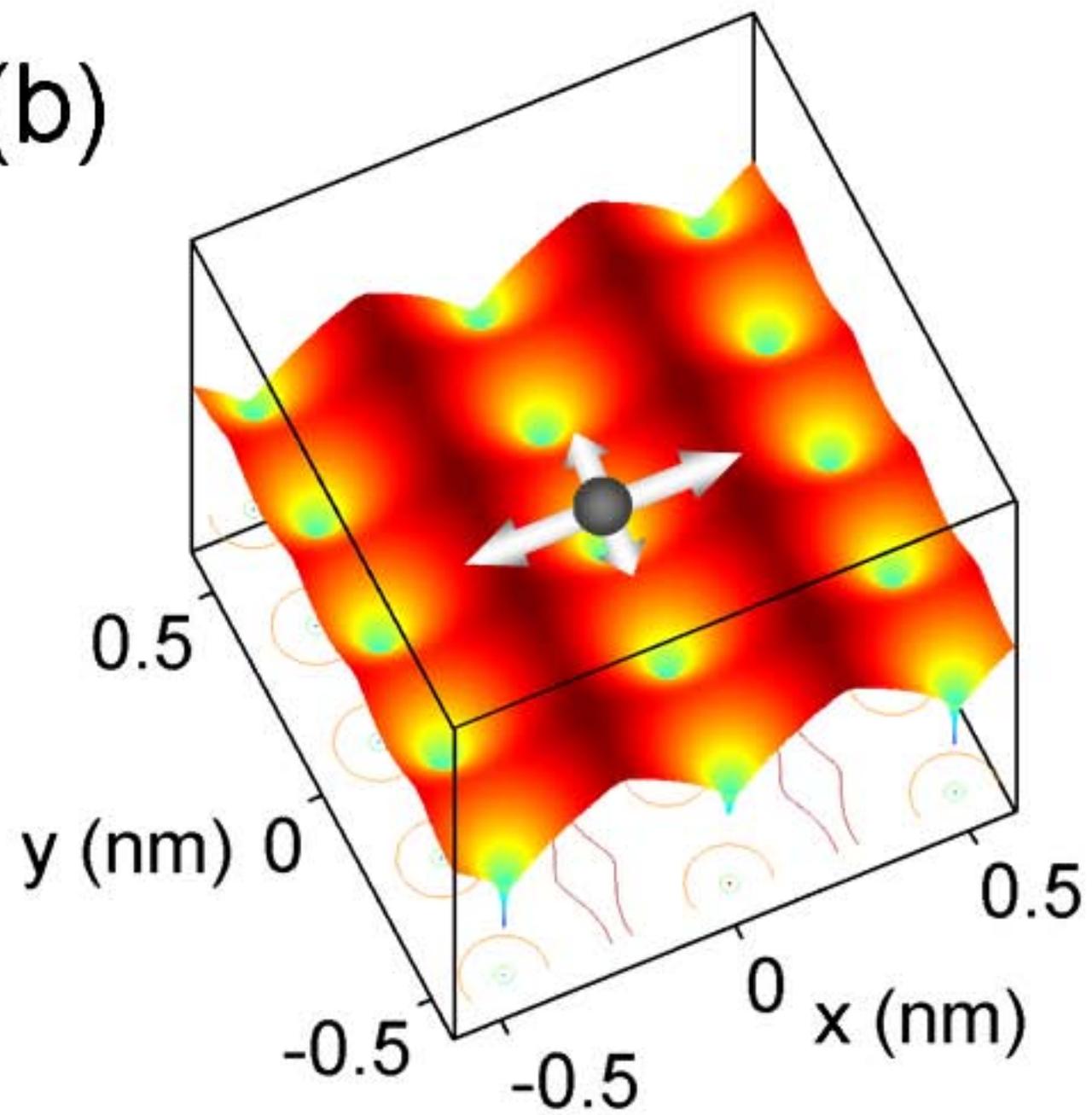

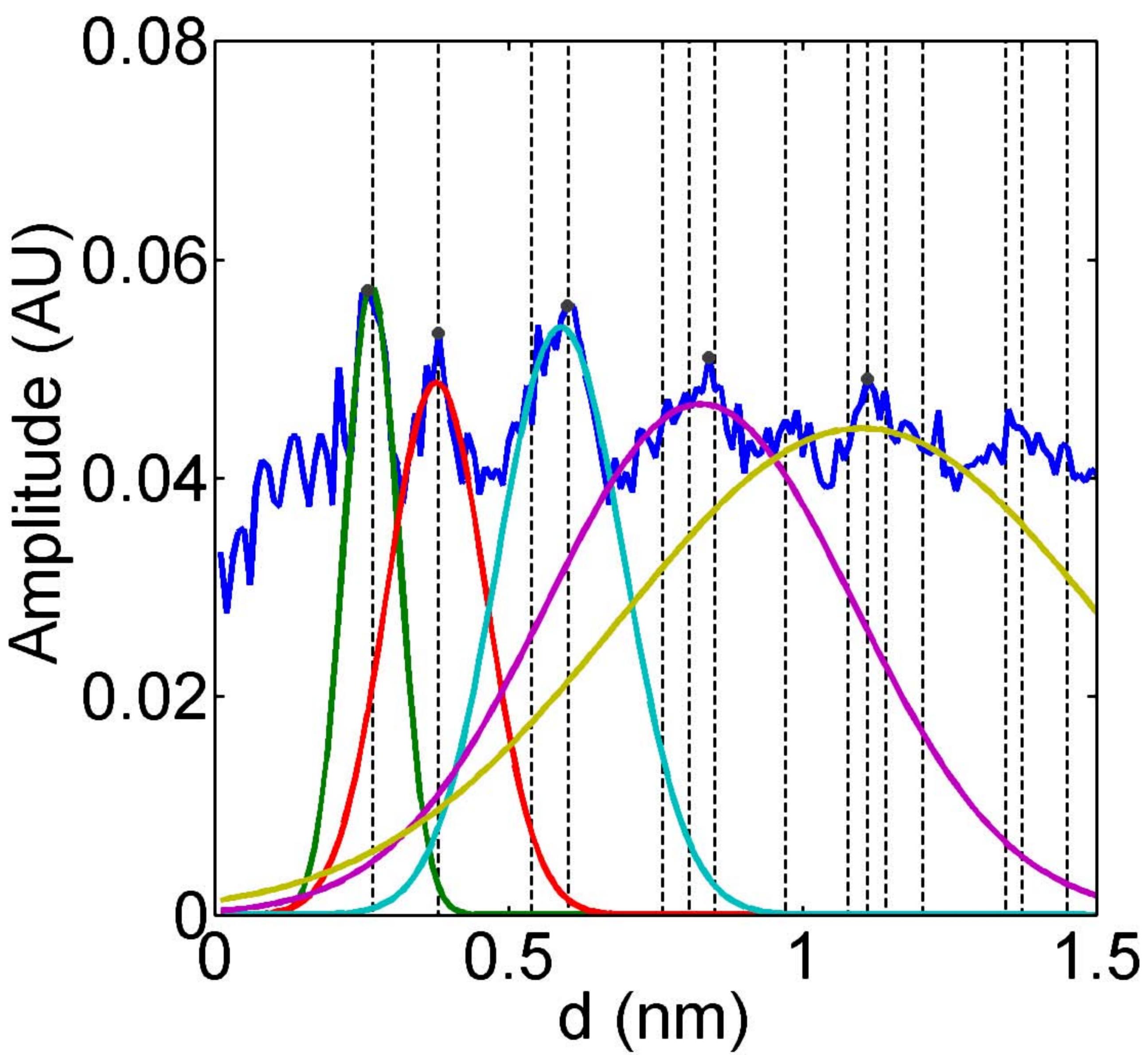

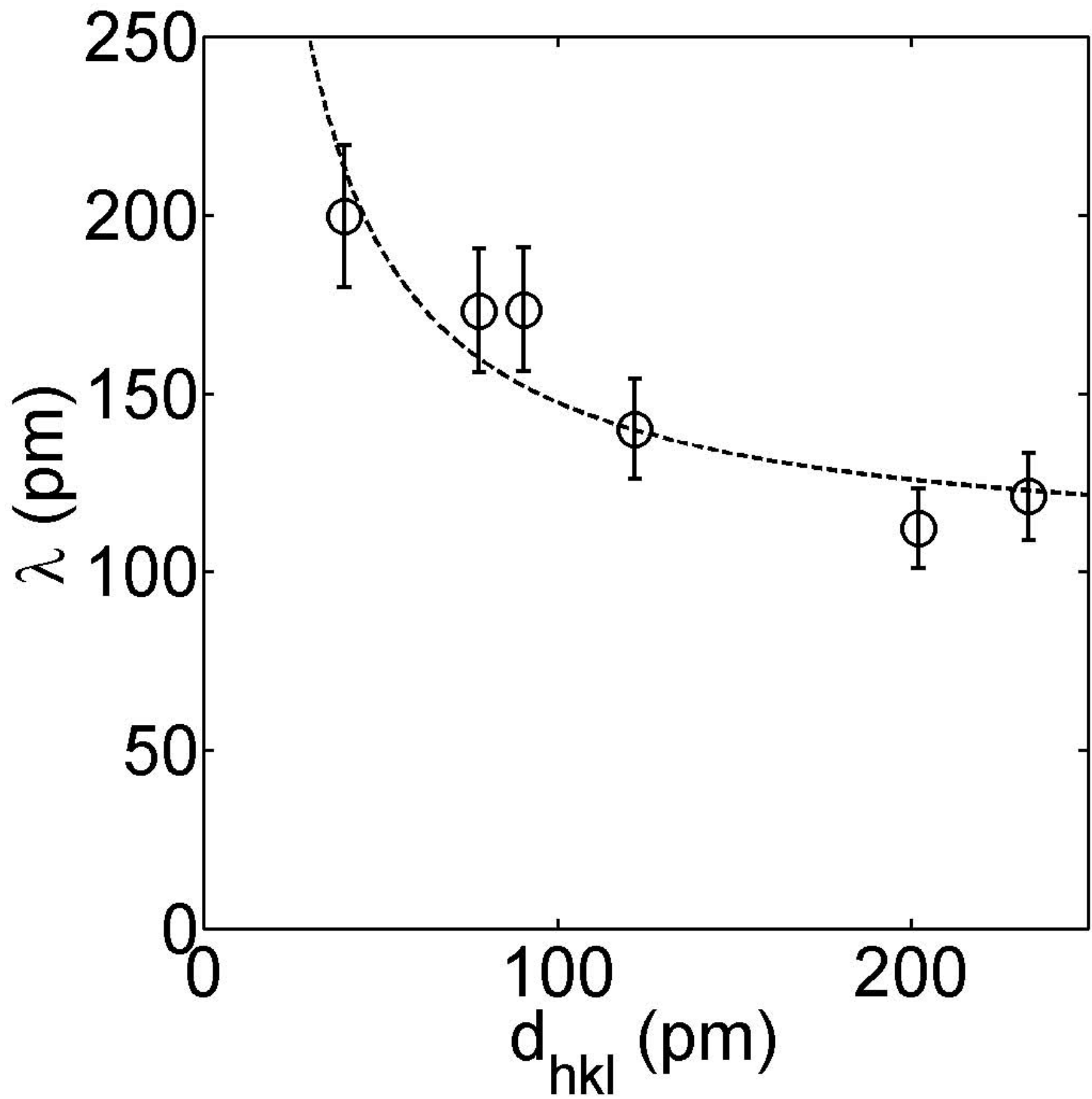

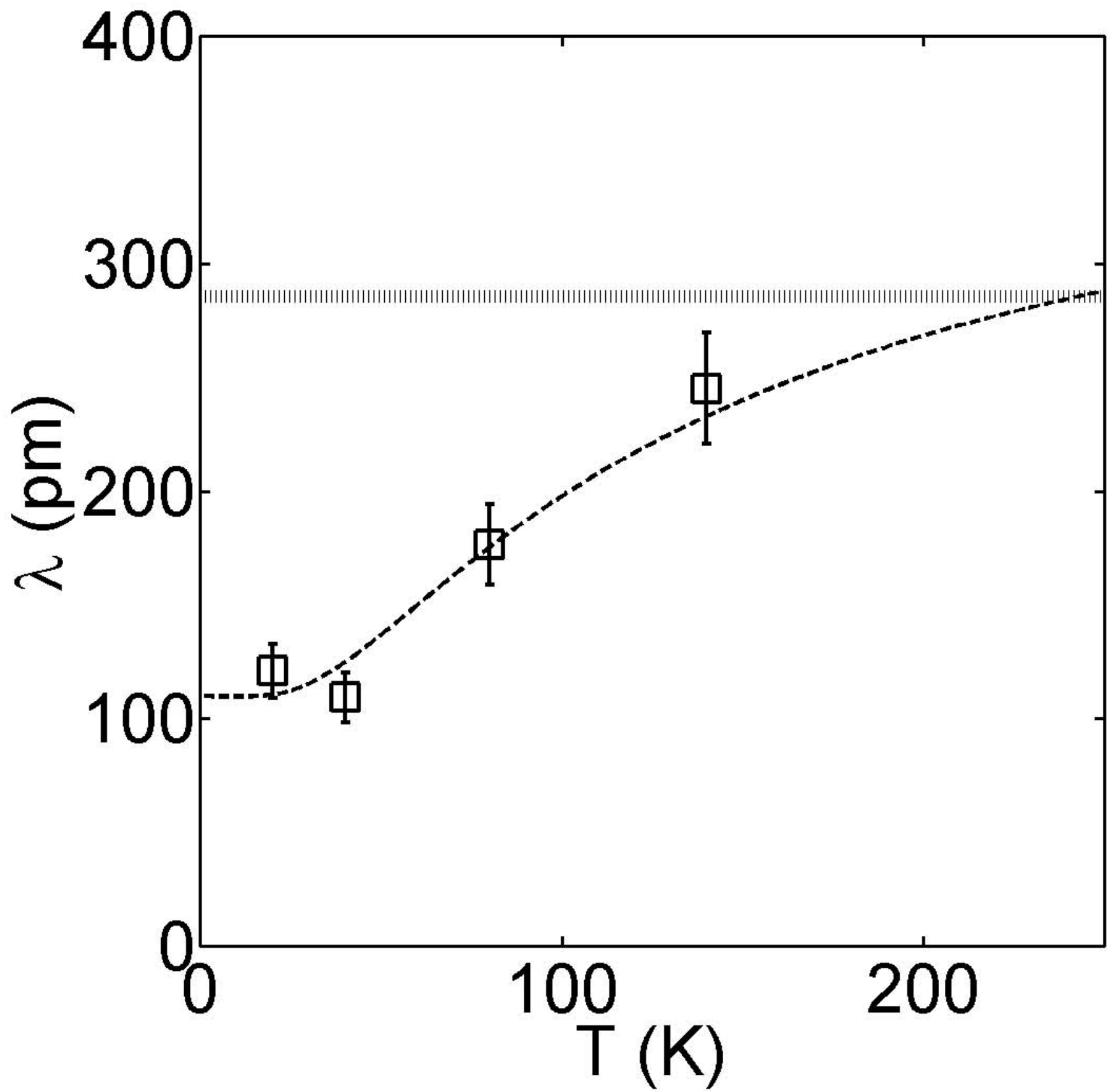

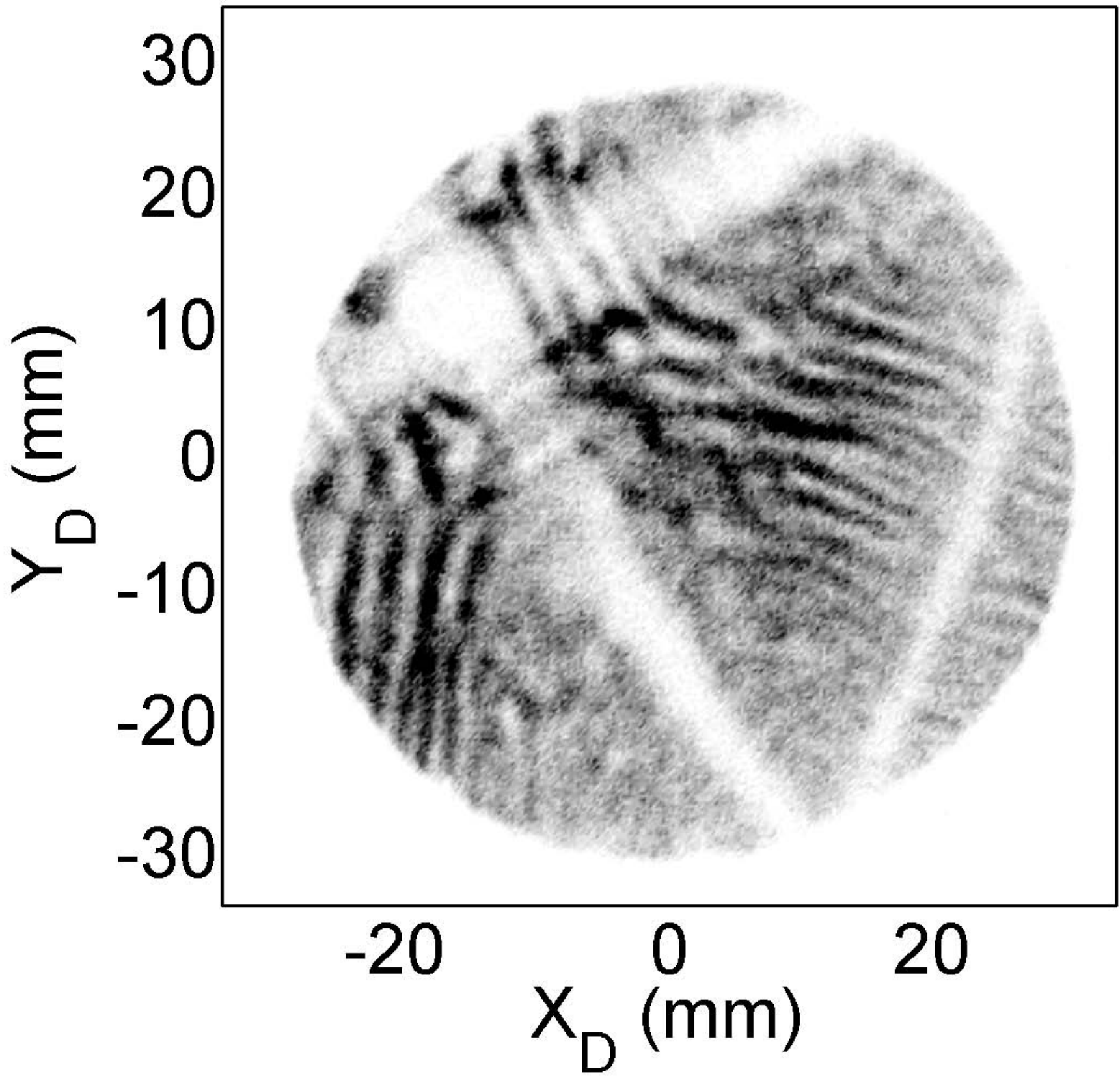